\documentclass[runningheads]{llncs}
\usepackage[usenames,dvipsnames]{xcolor} %for use in color links
\usepackage[utf8]{inputenc}
\usepackage{graphicx}
\usepackage{amsmath}
\usepackage{pdfpages}
\usepackage{booktabs}
\usepackage{caption}
\usepackage{threeparttable}
\usepackage{color}
\newcommand{\tmpframe}[1]{\fbox{#1}}
\renewcommand{\tmpframe}[1]{#1}
\begin{document}
\title{Segmentation of Defective Skulls from CT Data for Tissue Modelling}
%%%%%%%%%% MICCAI 2019 REVIEWS %%%%%%%%%%%%%%%
% https://docs.google.com/document/d/1sFTMxm1-s0gIgheT9nTwbXydeMqQczcB6ETG2Wr28nk/edit?usp=sharing
%%%%%%%%%%%%%%%%%%%%%%%%%%%%%%%%%%%%%%%%%%%%%%

\author{Oldřich Kodym\inst{1} \and Michal Španěl\inst{1,2} \and Adam Herout\inst{1}}
\authorrunning{O. Kodym et al.}
\institute{Graph@FIT, Brno University of Technology,\\
\email{ikodym@fit.vutbr.cz}
\and
TESCAN 3DIM, Brno, Czech Republic\\ 
\email{spanel@t3d.team}}

\maketitle              % typeset the header of the contribution
\begin{abstract}
In this work we present a method of automatic segmentation of defective skulls for custom cranial implant design and 3D printing purposes. Since such tissue models are usually required in patient cases with complex anatomical defects and variety of external objects present in the acquired data, most deep learning-based approaches fall short because it is not possible to create a sufficient training dataset that would encompass the spectrum of all possible structures. Because CNN segmentation experiments in this application domain have been so far limited to simple patch-based CNN architectures, we first show how the usage of the encoder-decoder architecture can substantially improve the segmentation accuracy. Then, we show how the number of segmentation artifacts, which usually require manual corrections, can be further reduced by adding a boundary term to CNN training and by globally optimizing the segmentation with graph-cut. Finally, we show that using the proposed method, 3D segmentation accurate enough for clinical application can be achieved with 2D CNN architectures as well as their 3D counterparts.

\end{abstract}
\section{Introduction}
Computer-assisted pre-surgical planning using generated 3D tissue models is seeing increasing use in personalized medicine~\cite{zille2018}. In the context of craniofacial surgery, the applications range from patient education, diagnosis and operative planning~\cite{urso1999} to patient-specific implant design~\cite{jardini2014}, mostly in the cranial area. The latter had been accelerated by the advent of additive manufacturing (AM), also known as 3D printing in recent years~\cite{mitsouras2015}. A typical workflow of producing a pre-surgical 3D tissue model consists of data acquisition, converting the data into patient model and optionally printing the model. Computed tomography (CT) is usually the modality of choice because of its unparalleled hard tissue contrast required for precise model shape extraction. As the manufacturing process is usually able to produce the model with a satisfactory precision, converting the raw CT data into an accurate patient model remains the most crucial step~\cite{martelli2016}. 

Precise segmentation of the patient skull is therefore critical. Although simple global thresholding followed by laborious post-processing and cleaning remains the most commonly used method in medical AM~\cite{eijnatten2018}, numerous semi- or fully automatic methods have been proposed for skull segmentation. Cuadros et al.~\cite{linares2018} used super-voxels followed by clustering and the level-set method has been applied to new-born skull segmentation in CT by Ghadimi et al.~\cite{ghadimi2016}. Following the success of the convolutional neural networks (CNN) in biomedical segmentation for both 2D~\cite{ronneberger2015} and 3D~\cite{cicek2016,milletari2016} settings, Minnema et al. used a simple patch-wise CNN for segmentation of skulls with defects for AM~\cite{minnema2018}. However, so far none of these methods have been able to show evidence that they are robust enough to be implemented into medical practice.

In this work, we propose an improved segmentation method that extracts region and boundary potentials using CNN and then uses graph-cut for globally optimal segmentation. The method outperforms methods based on conventional deep learning and other state-of-the-art methods of skull segmentation, and it produces results acceptable for the targeted use of 3D tissue modelling in the clinical practice. Furthermore, we directly compare 2D and 3D CNNs for segmentation and demonstrate that the benefit of using the 3D approach is not unequivocal.

\begin{figure}[t]
  \centering
  \tmpframe{\includegraphics[width=200px]{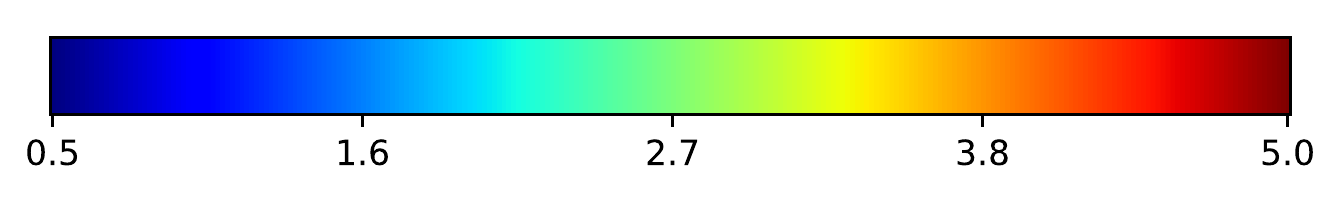}} 
  
  \tmpframe{\includegraphics[width=110px]{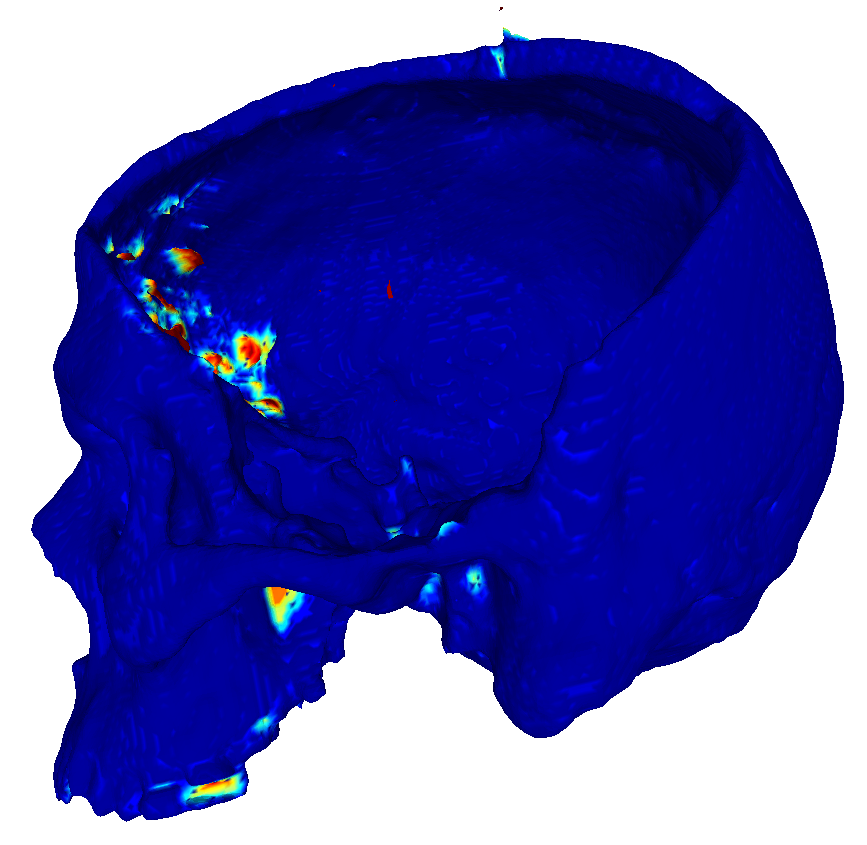}}
  \tmpframe{\includegraphics[width=110px]{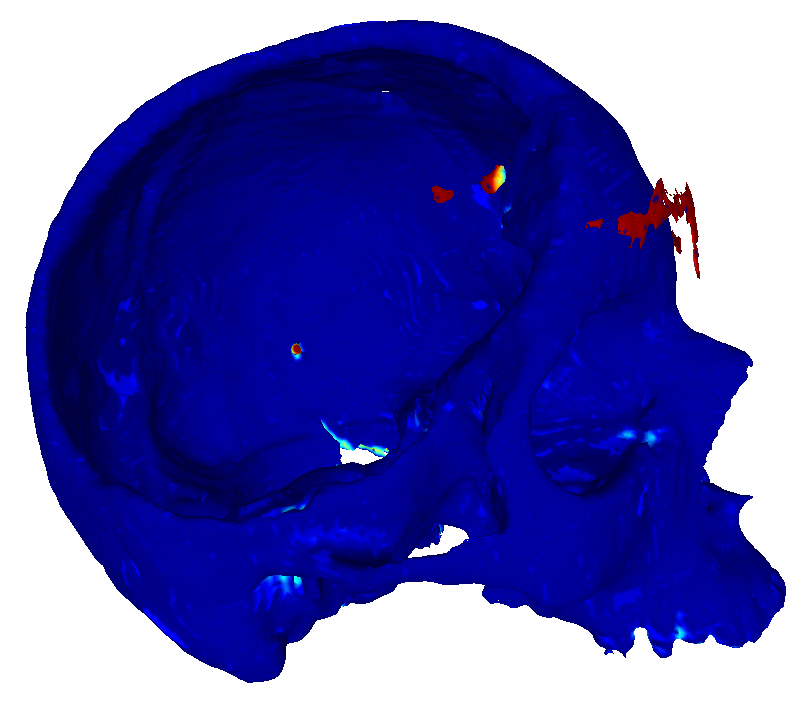}}
  \tmpframe{\includegraphics[width=110px]{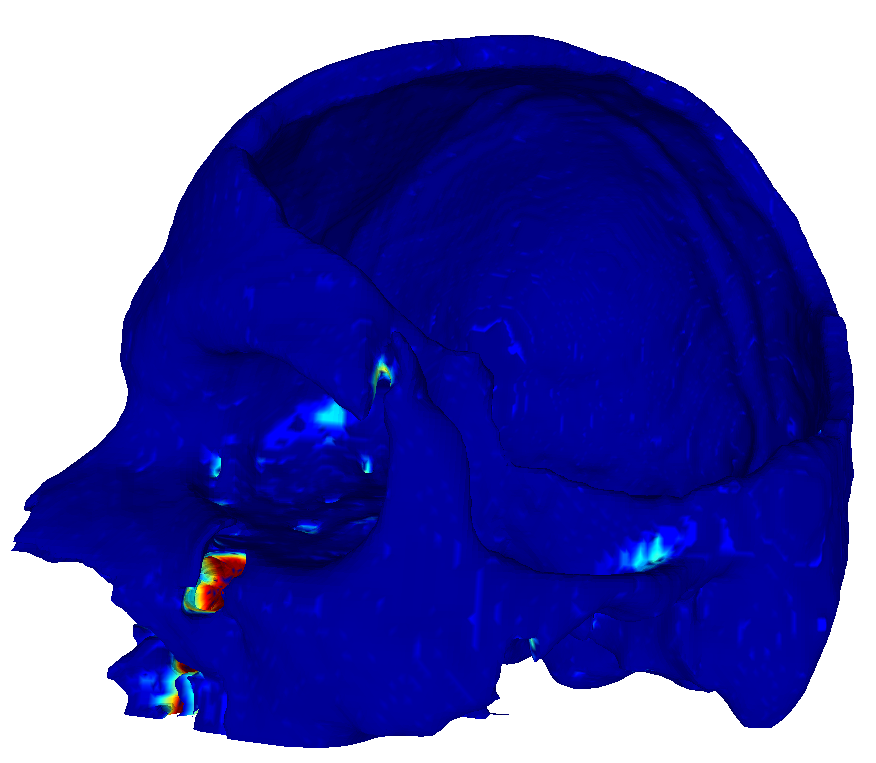}}
  \tmpframe{\includegraphics[width=110px]{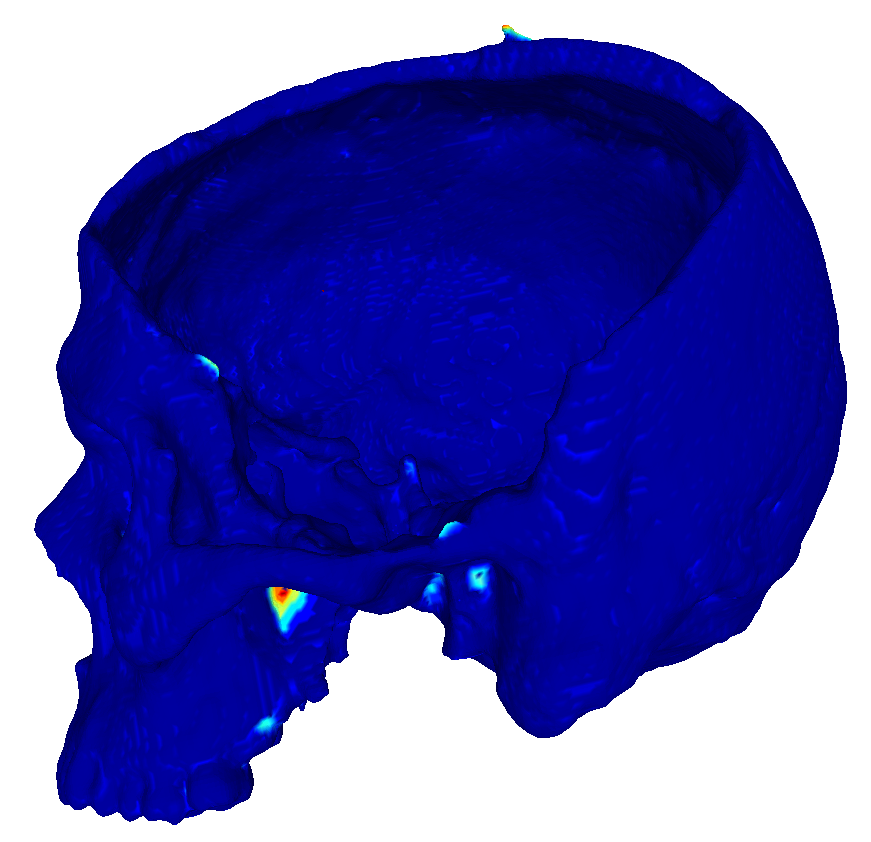}}
  \tmpframe{\includegraphics[width=110px]{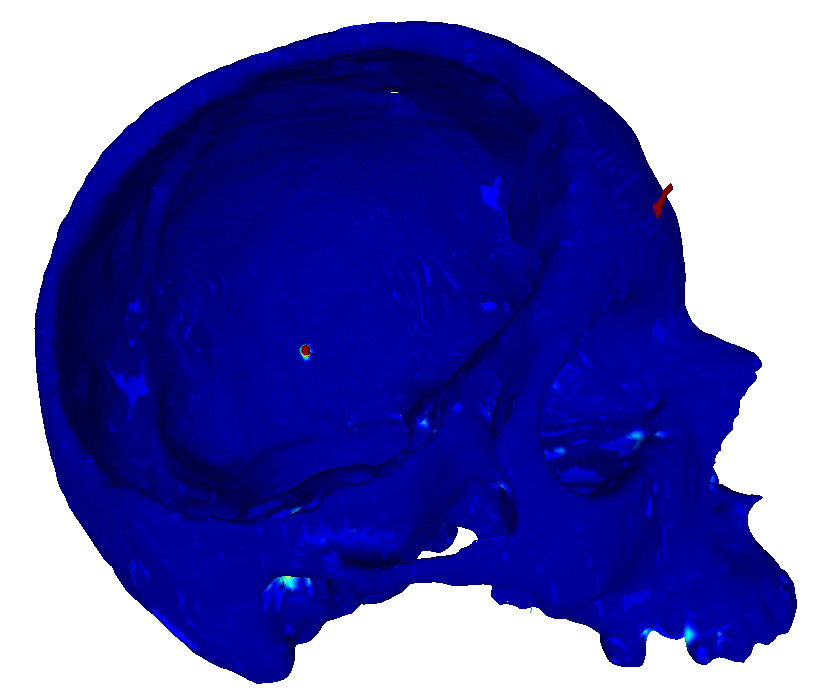}}
  \tmpframe{\includegraphics[width=110px]{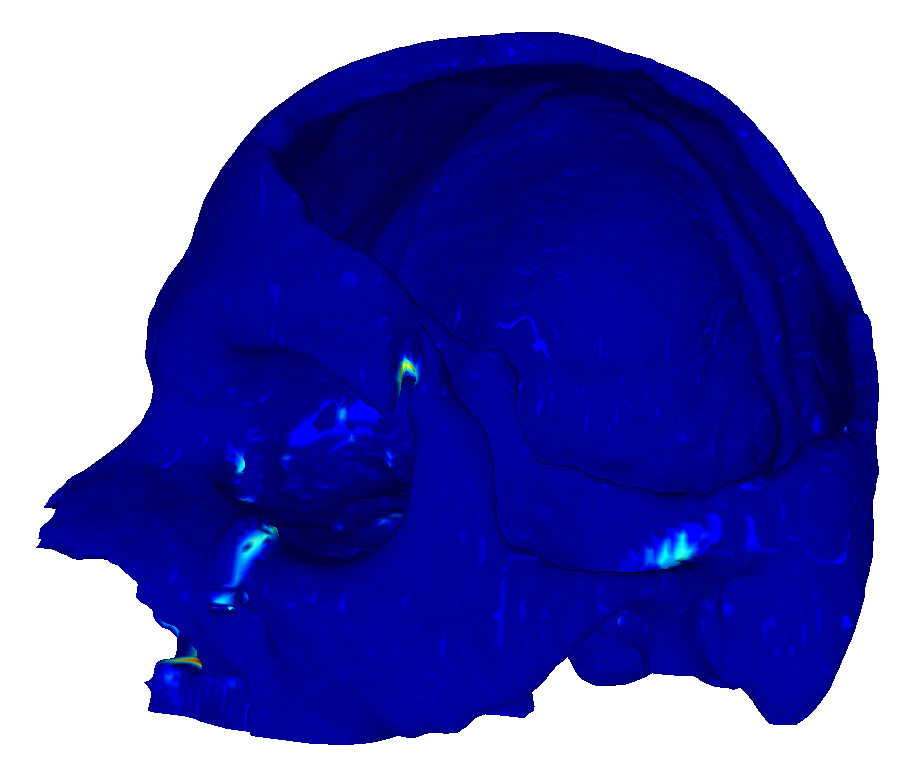}}
  
  \caption{Example renders of segmented skulls with the distance to the ground-truth surface in mm coded in color. Multi-view CNN segmentation outputs (top) and multi-view CutCNN segmentation outputs (bottom) are shown. To better display the differences, voxels with surface error of less than $0.5\,\mathrm{mm}$ are left dark blue.}
  \label{fig:render}
\end{figure}

%%%%%%%%%%%%%%%%%%%%%%%%%%%%%%%%%%%%%%%%%%%%%%%%%%
%%%%%%%%%%%%%%%%%%%%%%%%%%%%%%%%%%%%%%%%%%%%%%%%%%
%%%%%%%%%%%%%%%%%%%%%%%%%%%%%%%%%%%%%%%%%%%%%%%%%%
\section{Proposed Method}

We use the well known U-net model~\cite{ronneberger2015} as a baseline method for our segmentation experiments. We experimented with both multi-view (MV) ensemble of 3 orthogonal 2D U-nets as used in \cite{Chen2017} and fully 3D U-net \cite{cicek2016} since to authors' best knowledge, the current literature lacks direct comparison between the two approaches. The applied U-net slightly differs from the original architecture by using batch normalization and padding during convolutions, replacing the up-conv layers with bilinear up-sampling and reducing the initial number of convolutions to 16. The architecture of the 3D model is identical except that each convolution, max-pooling, and up-sampling operation is replaced by its 3D equivalent. The networks are trained until convergence using mini batches of shape \(24\times128\times128\) in case of 2D and \(4\times128\times128\times64\) in case of 3D model using the Dice loss function \cite{milletari2016}. 

\begin{figure}[t]
  \centering
  \tmpframe{\includegraphics[width=\textwidth]{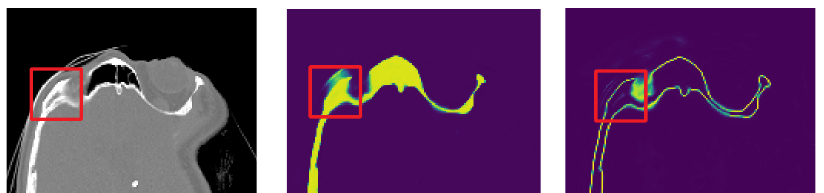}}
  \caption{Example CNN output slice, from left to right: Data, object probability map, edge probability map. Notice the segmentation error caused by an external object with density similar to that of the skull in upper left. The error is correctly separated by its detected edge.}
  \label{fig:PE}
\end{figure}

To improve segmentation performance on slightly out-of-distribution data (such as previously unseen medical material or defect shapes), we opted to apply 3D graph-cut segmentation on the CNN output. While this approach has been taken by other authors before \cite{lu2017}, we also modify our CNN model to output an edge probability for each voxel in addition to the object probability. Thus, the final layer of the CNN has 3 channels instead of the standard 2. Figure~\ref{fig:PE} illustrates how this step can provide additional boundary information to the graph-cut in comparison to simply using the conventional intensity or probability gradient. Another advantage of this approach is that since both region and boundary terms have similar dynamic range, finding optimal \(\lambda\) parameter of the graph-cut algorithm is simplified. We leave \(\lambda = 1\) throughout our experiments.

We train the network using the following form of the Dice loss function:
\begin{gather}\label{eq8}
\mathcal{L} =
    1-2\cdot\frac{\sum_{m=1}^{M}\left(p_{0}^{m}g_{0}^{m} + p_{1}^{m}g_{1}^{m} + p_{e}^{m}g_{e}^{m}\right)}
           {\sum_{m=1}^{M}\left(p_{0}^{m}+g_{0}^{m} + p_{1}^{m}+g_{1}^{m} + p_{e}^{m}+g_{e}^{m}\right)},
\end{gather}
where $p_0^m$ and $p_1^m$ are the probabilities of voxel $m$ belonging to class background and object respectively, and $g_0^m$ and $g_1^m$ are the corresponding ground-truth labels. Analogously, $p_e^m$ and $g_e^m$ are the probability and the ground-truth label of voxel belonging to the object edge. Edge map ground truth is obtained by subtracting the binary object from its morphologically dilated version, leaving a surface with single voxel thickness. Note that edge voxels overlap with the background voxels and the edge probability channel is therefore not included in the final softmax activation layer of the CNN.

Next, the output maps are converted into a 6-connected graph structure with the region terms $R\left(a\right)$ for voxel $a$ given by
\begin{gather}
\label{eq9}
R^{obj}(a)=-ln(p_1^a),\quad  R^{bkg}(a)=-ln(p_0^a)
\end{gather}
and the boundary term $B\left(a,b\right)$ between neighbouring voxels $a$ and $b$ given by
\begin{gather}
\label{eq9}
B(a,b) = -ln[max(p_e^a,p_e^b)].
\end{gather}

\begin{figure}[t]
  \centering
  \tmpframe{\includegraphics[width=\textwidth]{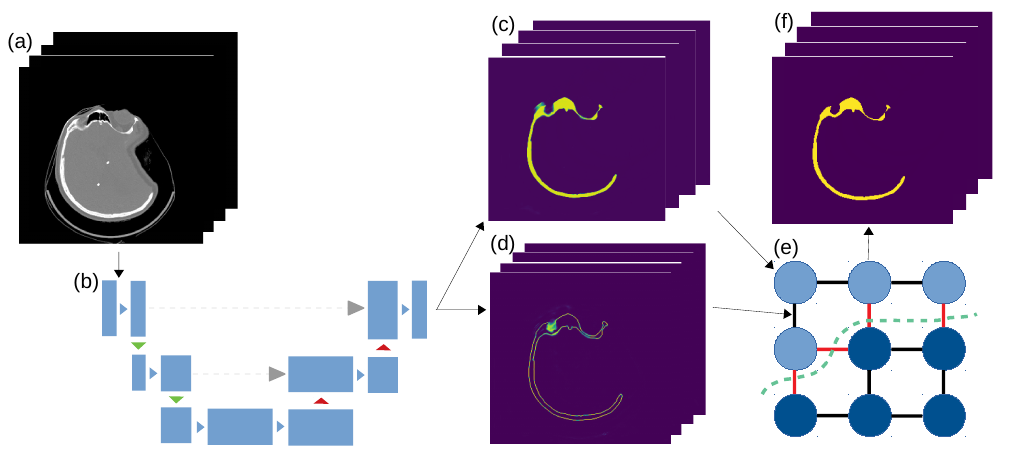}}
  \caption{Scheme of the proposed segmentation framework. Input data (a) are processed by a CNN model (b) to produce a probability map (c) and an edge strength map (d). These provide the boundary and region term for the graph-cut optimization step (e) which produces the binary output segmentation (f).}
  \label{fig:Method}
\end{figure}

Finally, globally optimal 3D segmentation can be obtained by finding minimum cut through this graph \cite{boykov2001}. This method will be referred to as CutCNN in the remaining parts of the paper. Note that while the CNN can be either MV (multi-view) or 3D, the graph-cut segmentation is always 3D. The method is summarized in~Figure~\ref{fig:Method}.

% \begin{figure}[t]
%   \centering
%   \tmpframe{\includegraphics[width=\textwidth]{architecture_ds.pdf}}
%   \caption{2D Segmentation model architecture. Series of two convolutions, batch normalizations and ReLU activations are used in each block. In 3D model each 2D operator (convolutions, max-pooling and bi-linear upsampling) is replaced with its 3D equivalent.}
%   \label{fig:CNNstructure}
% \end{figure}

%%%%%%%%%%%%%%%%%%%%%%%%%%%%%%%%%%%%%%%%%
%%%%%%%%%%%%%%%%%%%%%%%%%%%%%%%%%%%%%%%%%
%%%%%%%%%%%%%%%%%%%%%%%%%%%%%%%%%%%%%%%%%
\section{Experiments}
In this section, we present the skull tissue dataset on which the segmentation methods were evaluated. Then, we present the results of different segmentation methods on the dataset.

%%%%%%%%%%%%%%%%%%%%%%%%%%%%%%%%%%%%%%%%%
\subsection{Dataset}
Head CT scans of 199 different patients were available for this study. The scans were acquired for the purpose of patient skull modelling and its additive manufacturing or further patient-specific implant design. Therefore, pixel-wise ground-truth segmentation done by an experienced radiologist were also available for model training. The scans were acquired on multiple CT scanners using a variety of different acquisition protocols. The voxel size varied from \(0.38 \times 0.38 \times 0.38\)~mm to \(0.5 \times 0.5 \times 1.5\) mm. All volumes were re-sampled to isometric resolution of $1$ mm per voxel for the ablation experiments.

As the majority of these scans were acquired prior to a surgery, the skulls often contained various defects, fixation materials and other external objects. This makes fully automatic segmentation of these scans a challenging task, because many of these structures were only present in a single patient scan, making generalization difficult. 

\subsection{Metrics}
Multiple metrics were used to quantitatively compare outputs of different segmentation methods used in the study. Inspired by the MICCAI 2018 Medical Segmentation Decathlon challenge~\cite{MSD2018}, volumetric Dice coefficient and surface Dice coefficient were chosen. Furthermore, mean surface distance has been also included in the metrics as this is the recommended measure in area of medical tissue model preparation~\cite{eijnatten2018}. Implementations of the metrics used in this work are publicly available\footnote{\url{https://github.com/deepmind/surface-distance}}.

Dice coefficient (DC) is a well-known metric in medical segmentation domain. Given a number of true positive (TP) samples, false positive (FP) samples and false negative (FN) samples, the coefficient is given by

\begin{gather}\label{eq7}
DC = \frac{2 \cdot TP}{2 \cdot TP + FP + FN}.
\end{gather}

In case of volumetric Dice coefficient, number of voxels assigned an object label in output segmentation as well as in the ground-truth segmentation is used to compute TP while FP + FN correspond to the number of voxels assigned a different label.

To compute a surface Dice coefficient, the output and the ground-truth binary segmentation volumes are converted to polygon meshes. Each surface element in the output segmentation mesh is then considered a TP sample if the distance to the closest point on ground-truth surface is lower than threshold \(t\) and vice-versa. The surface elements in output and ground-truth meshes that do not fall under this threshold are counted as FN and FP, respectively. We chose the threshold to correspond to the voxel size in our experiment.

\subsection{Experimental Design and Results}
Performance of four different models has been evaluated in this study. Both 3D and MV CNN models and their CutCNN counterparts have been implemented in the TensorFlow framework. PyMaxflow library has been used for implementation of the graph-cut optimization. All experiments were run on a desktop system equipped with Nvidia Titan Xp GPU, an i5 intel core processor and 16GB RAM.

\begin{figure}[t]
  \centering
  \tmpframe{\includegraphics[width=250px]{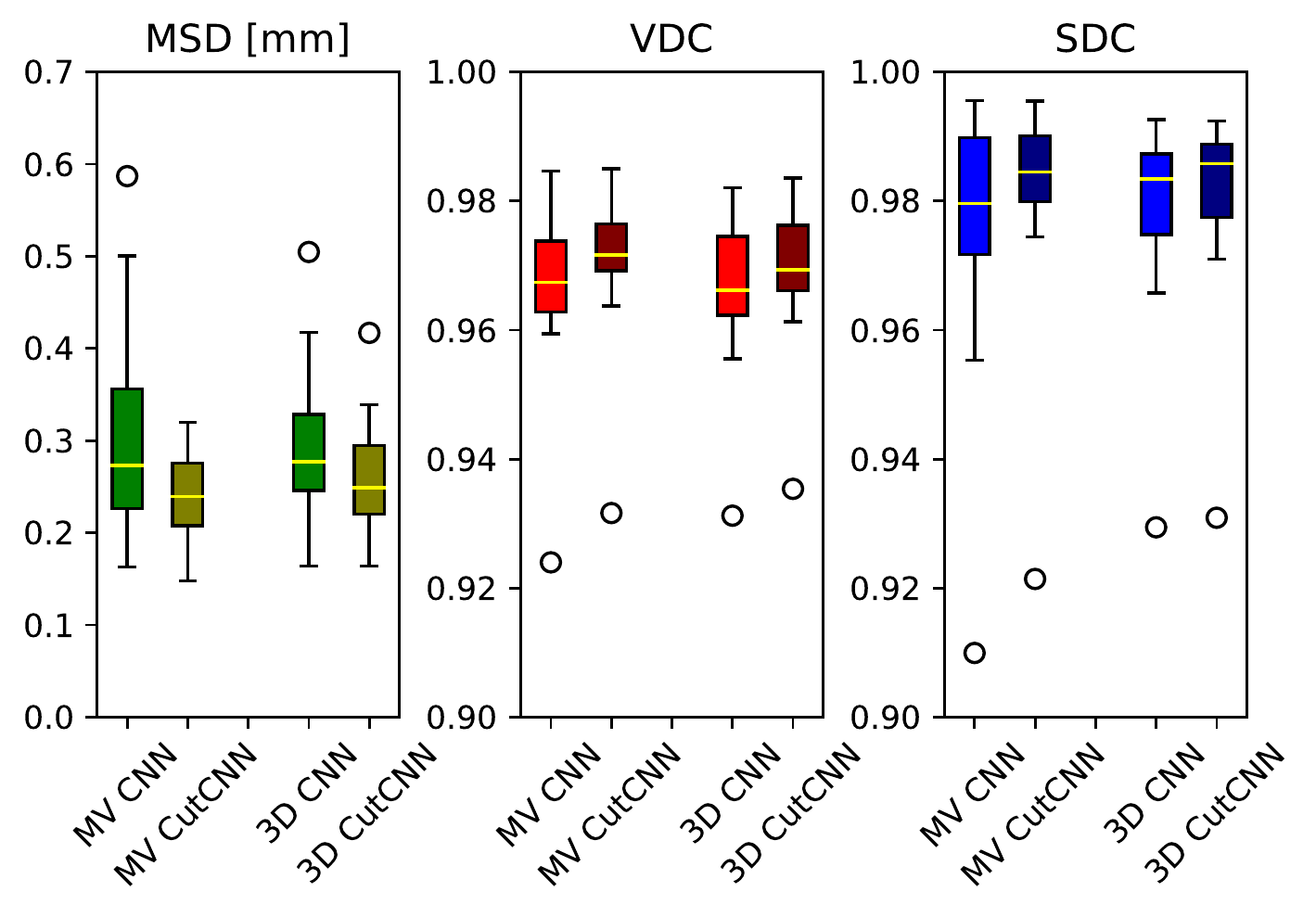}}
  \caption{Accuracy of standard multi-view (MV) and 3D CNN and their CutCNN counterparts. Results shown in terms of mean surface distance (MSD), volumetric Dice coefficient (VDC) and surface Dice coefficient (SDC).}
  \label{fig:graph}
\end{figure}

\begin{figure}[t]
  \centering
  \tmpframe{\includegraphics[width=\textwidth]{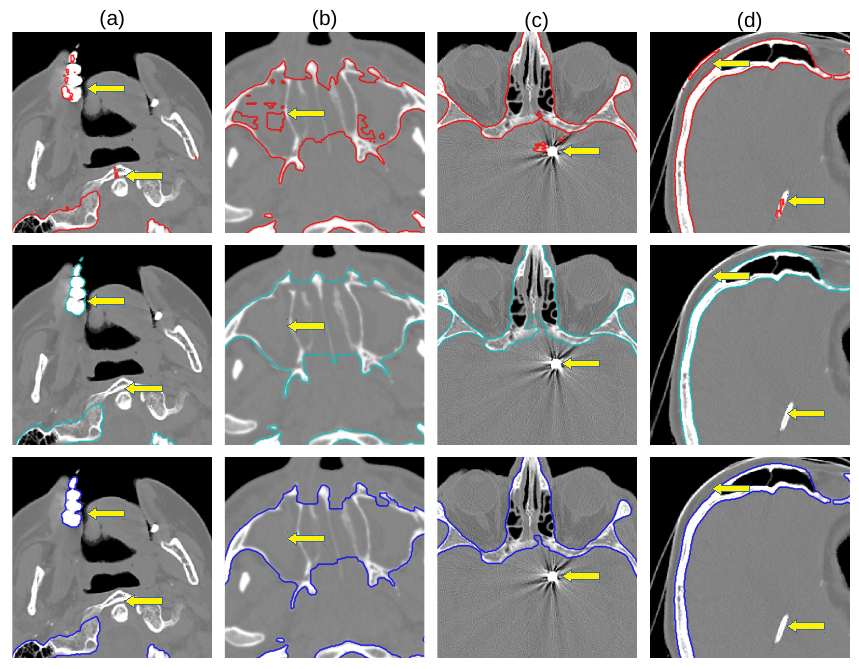}}
  \caption{Qualitative results shown for several chosen axial slices. From top to bottom: Multi-view CNN output (red), ground-truth (magenta), multi-view CutCNN output (blue).}
  \label{fig:more_slices}
\end{figure}

22 scans were randomly selected as test subjects for the experiment, leaving 177 skulls for model training. Using convolutional kernels of size 3 in all the CNN models results in the 3D model having the same number of trainable parameters as the sum of the three orthogonal 2D models. The comparison 
between the MV ensemble and the 3D approach can therefore be considered an ablation study to an extent. CutCNN models also have a similar number of parameters, the only difference being the final edge probability output layer. Quantitative comparison of results of each method are presented in Figure \ref{fig:graph} and Table \ref{tab:Comparison_dice}. Further qualitative results are shown in Figure \ref{fig:more_slices} and \ref{fig:render}. 

\begin{table}[t]
  \centering
  \caption{Comparison of segmentation methods using mean surface distance (MSD) [mm], volumetric Dice coefficient (VDC) and surface Dice coefficient (SDC).}
  \label{tab:Comparison_dice}
  \setlength{\tabcolsep}{3mm}
  \begin{tabular}{lccccccc}
    \midrule[0.5pt]
      \textbf{Method} & \textbf{MSD} & \textbf{VDC} & \textbf{SDC}\\
      \midrule[0.5pt]
      MV CNN & 0.37 & 96.7 & 97.1\\
      3D CNN & 0.35 & 96.7 & 97.0\\
      MV CutCNN & \textbf{0.31} & 97.7 & \textbf{98.3}\\
      3D CutCNN & 0.32 & \textbf{98.0} & 98.1\\
      \midrule[0.5pt]
     * Minnema et al. \cite{minnema2018} & 0.44 & 92.0 & -\\
      * Linares et al. \cite{linares2018} & - & 91.5 & -\\
      \midrule[0.5pt]
    % \bottomrule
  \end{tabular}
  \begin{tablenotes}
      \small
      \centering\item* Results obtained on different datasets
    \end{tablenotes}
\end{table}

\section{Discussion}
CutCNN segmentation framework resulted in a performance gain in all cases in terms of every metric used in the experiment over standard CNN approaches. The output of CNN object probability map often contains errors near external objects or smaller tissue defects as these are scarce in the training data distribution. However, the graph-cut optimization guides the resulting binary segmentation towards a spatially consistent and compact shape, often eliminating these artifacts if a detected edge corresponds mostly to the correct object boundary. This effect is further illustrated in Fig.~\ref{fig:render}.

Our second observation is that using 3D convolutional kernels has a rather small effect on the final segmentation precision quantitatively compared to the MV approach. However, although the quantitative difference is small, for applications in medical additive manufacturing, it is important to avoid ragged segmentation output which may result from MV CNN in areas of lower model certainty. These include for example teeth, which are challenging to detect, especially when the lower and upper teeth are in contact (see Figure~\ref{fig:more_slices}~a), or maxillary sinus, which is often enclosed in order to improve mechanical stability of the manufactured model~(see Figure~\ref{fig:more_slices}~b). Therefore, 3D U-nets are often considered necessary to avoid these discontinuities caused by slice-by-slice processing.

However, this artifact can also be addressed by employing the CutCNN framework since ragged segmentation boundary introduces a high boundary-term penalization during optimization and it is therefore avoided in the final binary segmentation. Thus, employing CutCNN allows the decision between 3D or multi-view approach to be merely a technical choice. Using 2D models can offer some advantages, such as faster training of deeper models with less overfitting~\cite{Chen2017}.

We also evaluate the performance of the proposed method in the context of existing related work in skull segmentation. In terms of volumetric Dice coefficient, the proposed method achieved performance of $0.977\pm0.019$ in the multi-view scenario and $0.980\pm 0.013$ in the 3D scenario. This result is considerably higher than that of $0.92\pm 0.04$ reported by Minnema et al.~\cite{minnema2018}. This is probably caused by the lower resolution in our experiments and by several limiting factors in the other works, including the small training set that only allowed for a smaller CNN architecture and employing a patch-based approach. To our best knowledge, the presented work is the first to apply a fully automatic segmentation approach to a pathological skull dataset of this size. Furthermore, we also achieve a low mean surface distance with the proposed method, namely $0.31\pm 0.33$ mm. 

We also trained the multi-view CutCNN model with isometric resolution of $0.5$ mm per voxel to facilitate enough precision for clinical practice with almost no loss in accuracy. Preliminary testing of the proposed method by experts in medical tissue modelling practice showed that the results of this model are accurate enough to substantially reduce the amount of time spent by creating the model in practice when compared to currently used semi-automatic segmentation methods.

\section{Conclusions}
In this work, we presented CutCNN, an improved hard tissue segmentation method which integrates the CNN output with graph-cut segmentation. The results of such a method surpassed the commonly used CNN architectures such as 3D and multi-view U-nets as well as other competitive methods in the skull segmentation domain. The object and edge probability maps in combination with graph cut provide a compact and smooth final tissue segmentation while adding very little computational cost. This method could therefore be used to improve the performance of any semantic segmentation task given that the edges are well defined in the data. In the future, to deal with any remaining segmentation errors, user interaction can be introduced to the method on both CNN and graph-cut level as the output of both steps can be improved through user scribbles in an iterative fashion. This will further reduce the time spent producing accurate tissue model.

\vspace{8pt}
\noindent{\small \textbf{Acknowledgements.} This work was supported in part by the company TESCAN 3DIM. We also gratefully acknowledge the support of the NVIDIA Corporation with the donation of one NVIDIA TITAN Xp GPU for this research.}


\end{document}